\documentclass[%
 reprint,
superscriptaddress,
amsmath,amssymb,
prb,
]{revtex4-1}

\usepackage{graphicx}% Include figure files
\usepackage{dcolumn}% Align table columns on decimal point
\usepackage{bm}% bold math
\usepackage{amsmath}
\usepackage{nccmath} %for shortskipping to reduce spaces between equations

\newcommand*{\citen}[1]{%
  \begingroup
    \romannumeral-`\x % remove space at the beginning of \setcitestyle
    \setcitestyle{numbers}%
    \cite{#1}%
  \endgroup   
}

\DeclareMathAlphabet\mathbfcal{OMS}{cmsy}{b}{n}  % for bold mathcal symbs

\begin{document}

\preprint{APS/123-QED}

\title{Fully characterized linear magnetoelectric response of 2D monolayers from high-throughput first-principles calculations}

\author{John Mangeri}
 \email{johnma@dtu.dk}
 \author{Thomas Olsen}
 \email{tolsen@fysik.dtu.dk}
\affiliation{%
Computational Atomic-scale Materials Design (CAMD), Department of Physics, Technical University of Denmark, \\ 
 Anker Engelunds Vej 101 2800 Kongens Lyngby, Denmark\\
}%
\date{\today}%

\begin{abstract}
We screen 4784 stable monolayers from the Computational 2D Materials Database (C2DB) and identify 57 ferromagnetic (FM) and 67 antiferromagnetic (AFM) compounds that should exhibit linear magnetoelectric (ME) effects.
Using density functional theory, we compute contributions from the spin and orbital angular momentum as well as lattice-mediated and clamped-ion analogs to fully characterize the linear ME tensor in the static limit.
We observe a general trend that AFM ordering gives rise to a larger ME response compared to FM ordered monolayers.
Using a typical van der Waals interlayer distance, we find that AFM $\mathrm{Mn}_2\mathrm{SI}_2$ exhibits the strongest component of linear ME response, providing approximately 580 ps/m.
This is two orders of magnitude greater than in prototypical $\mathrm{Cr}_2\mathrm{O}_3$ but comparable to the largest ME response measured in bulk $\mathrm{TbPO}_4$ (280-740 ps/m).
We also search for antimagnetoelectricity and find a number of FM and AFM compounds with antiferroic tensor entries.
By demonstration of select examples and analysis of our full data set, we argue that inclusion of all contributions (spin, orbital, lattice-mediated and clamped-ion) is of crucial importance for reliable predictions of the total ME response.
\end{abstract}


\maketitle

%\tableofcontents

\section{\label{sec:intro} Introduction}

The paradigm of computational materials discovery continues to accelerate the search for novel structures and their properties\cite{Jain2013, Saal2013, Gjerding2021}.
A particular interesting class of compounds are those with simultaneously broken inversion and time-reversal symmetry or those where the magnetic ordering breaks the inversion symmetry of the underlying nuclear positions (the so-called $\mathbfcal{PT}$ symmetry).
Both symmetries underpin a coupling between the electric and magnetic dipole order, and a nontrivial response to the electromagnetic field emerges and gives rise to the magnetoelectric (ME) effect\cite{Rivera1994, Weiglhofer2003, Fiebig2016, Spaldin2019}.
If the ME effect is strong, as is the case in multiferroics, it is favorable for the advancement of spintronic-based technologies\cite{Fusil2014, Manipatruni2019, Huang2024}.
In addition, two-dimensional (2D) crystals are a very promising platform, as they can be suitable for scaling down to a few atomic layers \cite{Schwierz2010, Ferrari2015, Novoselov2016, Jiang2018, Gjerding2021}.
Some 2D MEs have been found and various interesting properties have been experimentally reported\cite{Kurumaji2013, Kim2018, Jiang2018, Song2022, Selter2023, Wang2023b, Aoki2024, Gao2024, Zhang2025}.
%
% Jiang2018 is bilayer CrI3
% Kurumaji2013 is NiI2,CoI2
% Kim2018 is NiPS3
% Song2022, Gao2024 are Nature papers on NiI2
% Zhang2025, review on Ni2, NiPS3, Fe3GeTe2, and Fe5GeTe2 
%    the last two are not multiferroic but display some interesting current-control of magnetism
% Selter2023, Wang2023b is CuCrP2S6 (Cu2Cr2P4S12 in our dataset)
%
% J: I believe all but NiPS3 are metallic
%
%
However, ME compounds are rather scarce and only relatively few specific materials have been reported.
We refer to the comprehensive list of experimentally known MEs provided by Watanabe and Yanase\cite{Watanabe2018}, which shows that there is a clear need to implement theoretical and data-driven approaches to find more of these compounds.
The linear magnetoelectric effect may be quantified by the tensor
\begin{align}\label{eq:linear_ME}
    \alpha_{ij} = \mu_0 \left(\frac{\partial M_i}{\partial \mathcal{E}_j}\right)\bigg|_{\mathbfcal{H}} = \left(\frac{\partial P_j}{\partial \mathcal{H}_i}\right)\bigg|_{\mathbfcal{E}}.
\end{align}
The vectors $\mathbf{M}$ and $\mathbf{P}$ are, respectively, the bulk magnetization and electric polarization of the crystal, \emph{induced} by applied homogeneous and  \emph{static} electric $(\mathbfcal{E})$ and magnetic $(\mathbfcal{H})$ fields. 
Since Eq.~(\ref{eq:linear_ME}) changes sign under $\mathbfcal{T}$ or $\mathbfcal{P}$, a non-vanishing linear ME response requires the absence of both these symmetries.
However, the combined symmetry $\mathbfcal{PT}$ renders the linear ME tensor invariant and may or may not be present in linear magnetoelectrics.
Many experimental investigations have been initiated\cite{Vaknin2004, Mufti2011, Saha2016, Ghara2017, Yanda2019, Liu2021, Shahee2023, Fogh2023a, Fogh2023b,  Du2023}, particularly in insulating antiferromagnets (AFMs), to probe the strengths of ME coupling, finding a great deal of microscopic complexity even in the linear limit.
As such, there is a need for theoretical predictions to help quantitatively understand this effect.
The lattice-mediated (LM) contribution to the linear ME response was discussed in a seminal work by \'{I}\~{n}iguez \cite{Iniguez2008}.
Here, distortion of the nuclear coordinates under the field and the subsequent rearrangement of the electronic degrees of freedom on the perturbed Born-Oppenheimer energy surface drive the LM contribution to ${\bm \alpha}$.
Following this development, Bousquet \emph{et} \emph{al} \cite{Bousquet2011} showed that the magnetic field-induced polarization of the electron cloud around fixed atomic positions comprises a clamped-ion (CI) contribution to ${\bm \alpha}$.
Both reports provided a methodological pathway to quantitatively resolve contributions to ${\bm \alpha}$ using first-principles methods.
In the linear regime, the two distinct effects are simply additive such that ${\bm \alpha} = {\bm \alpha}^\mathrm{LM}+{\bm \alpha}^\mathrm{CI}$.\cite{Malashevich2012, Ricci2016, Mangeri2024}.
Since the total momentum of the electron carries both a spin (S) and orbital (L) part, the LM and CI contributions split into four distinct contributions\cite{Malashevich2012, Mangeri2024}:
\begin{align}\label{eq:total_alpha}
{\bm \alpha} = {\bm \alpha}^\mathrm{LM,S} + {\bm \alpha}^\mathrm{CI,S} + {\bm \alpha}^\mathrm{LM,L} + {\bm \alpha}^\mathrm{CI,L}.
\end{align}
Density functional theory (DFT) calculations reported by Malashevich \emph{et al}\cite{Malashevich2012} provided a \emph{full characterization} of ${\bm \alpha}$ connecting the relative strengths and signs of the LM and CI (as well as spin and orbital magnetization) effects in the prototypical AFM insulator $\mathrm{Cr}_2\mathrm{O}_3$.
Here, the spin order imposed on the atomic coordinates is $\mathbfcal{PT}$-symmetric providing a nonzero Eq.~(\ref{eq:linear_ME})\cite{Dzyaloshinskii1959, Dzyaloshinskii1960}.
In that work, it was shown that CI effects in the transverse components (with respect to the spin) of ${\bm \alpha}$ can constitute almost $30\%$ of the total value.
The theoretical description of the ME in $\mathrm{Cr}_2\mathrm{O}_3$ continues to be studied as recently as in Ref.~[\citen{Bousquet2024}] investigating how different numerical/DFT approaches perform in evaluating ${\bm \alpha}$.
Thus, $\mathrm{Cr}_2\mathrm{O}_3$ is established as a useful benchmark material platform for theoretical studies of the linear ME effect.
Experiments measuring non-reciprocal directional birefringence\cite{Krichevtsov1993, Krichevtsov1996} in $\mathrm{Cr}_2\mathrm{O}_3$, inferred that the strengths of the optical ${\bm \alpha(\omega)}$ can be comparable to the static counterpart, suggesting that evaluations from first-principles, in the $\omega \to 0$ limit, are useful for understanding this phenomenon.
%

%
%
%See Refs. by H. Schmid for an older historical record - still comparable order to CRO with the exception of TbPO4. 
%
%         Vaknin2004 (LiNiPO4)
%         Saha2016: spinel Co3O4 and Mn(Al,Ga)2O4
%         Ghara2017: spinel CoAl2O4
%         Mufti2011: MnTiO3
%         Shahee2023: Cu3TeO6 (1.8 ps/m)
%         Fogh2023a,b: Li(Fe,Ni)PO4
%         Yanda2019: Sm2BaCuO5, Pnma
%

%
Even if the ME is allowed by symmetry, there is no simple way to \emph{a} \emph{priori} determine whether or not the effect is strong.
In particular, experimental measurements of the linear ME effect in many compounds have only found a weak or comparable coupling to that of $\mathrm{Cr}_2\mathrm{O}_3$\cite{Vaknin2004, Mufti2011, Saha2016, Ghara2017, Yanda2019, Shahee2023, Fogh2023a, Fogh2023b}.
Several physical factors could play a role.
In particular, the magnetic network geometry, the magnitude and sign of exchange interactions, the magnetic anisotropy, the electronic band gap, spin-orbit coupling (SOC), as well as neighboring ligand environments, and the total (spin plus orbital) magnetization of the magnetic element could \emph{all} influence the response in the static limit.
Therefore, data-driven approaches are needed to uncover trends and identify or assist in the design and discovery of new ME materials.
Such studies will also help to fully characterize the microscopic mechanisms that drive the effect, thus stimulating next-generation technological innovation.
In the present work, we perform a high-througput search for ME two-dimensional materials (MLs) based on the Computational Two-dimensional Materials Database (C2DB).
We identify 124 2D magnetoelectrics, several of which are known as bulk van der Waals bonded compounds and some of which have been exfoliated down to the few layer limit\cite{Carretta2002,  Hugonin2008, Baithi2023, Peng2020, Lai2019, Selter2023, Wang2023b, Chernoukhov2024, Lu2024}.
%
% Carretta2002 VOMoO4 (Mo2V2O10 in our dataset )
% Hugonin2008 CoSb2Br2O3 (Co2Sb4O6Br4 in our dataset)
% Baithi2023  CrPSe3 (Cr2P2Se6 in our dataset)
% Peng2020 AgVP2Se6
% Lai2019, Selter2023 CuCrP2S6 (Cu2Cr2P4S12 in our dataset)
% Chernoukhov2024 (Al2Mn2Se5 in our dataset)
% Lu2024, Cr2Se3
%
For all materials, we perform a series of DFT calculations to fully characterize all 9 components of ${\bm \alpha}$ as well as the 4 independent contributions given by Eq.~(\ref{eq:total_alpha}).
Our data reveal a number of strongly ME materials,
and we can classify these as spin- or orbital-driven, and designate the principal contributions as either lattice-mediated or purely electron-mediated.
From all the 2D magnetoelectrics, we find that the AFM ML
$\mathrm{Mn}_2\mathrm{I}_2\mathrm{S}$ hosts the largest response.
Using a typical van der Waals interlayer distance, the response can be converted to a 3D bulk value of $|\alpha|\sim 580$ ps/m.
This is primarily driven by the lattice-mediated spin effect and is comparable to the largest linear ${\bm \alpha}$ measured in $\mathrm{TbPO}_4$\cite{Rivera2009, Grams2024}.
The LM term is typically considered to be the main contribution to the static ME response, due to the large responses induced by ionic motion in a field\cite{Bousquet2011}.
%
% Here, we pick out the statement from Bousquet2011 to back the claim 
%   "However, this approach computes only the so-called ‘‘lattice-mediated’’ part of and ignores purely electronic contributions to the response. The common justification is that such contributions are expected to be weak, just as in strong dielectrics the electronic response is negligible compared to the ionic contribution"
%
Similarly, the orbital contribution is expected to be weak because many materials display quenched orbital moments.
However, our results demonstrate that this is not the case in general.
We find that only 21$\%$ of the monolayers exhibit a dominant LM spin term, whereas a larger percentage (26$\%$) have the majority contribution coming from the orbital magnetization through the LM effect.
The CI spin and CI orbital contributions that constitute the principal part of the ME response, in the remaining part of the materials, are 40$\%$ and 13$\%$ respectively.%

We observe that the ME response is generally stronger in AFMs as compared to ferromagnets (FMs).
We also propose that one cannot neglect any of the contributions when characterizing this ME effect from first-principles calculations.
We exemplify, in a targeted example, that the total ME response can almost vanish when considering all of the contributions due to opposing relative signs treated on equal footing.
In other cases, the ME effect can be dominated by a single contribution, which may be either of the four contributions discussed above.

Finally, we discuss the antimagnetoelectric effect, a type of hidden response recently proposed\cite{Verbeek2023}, and highlight both FM and AFM compounds that exhibit it.
The paper is organized as follows: In Secs.~\ref{sec:screening_comp_details} and \ref{sec:DFT_details}, we describe the computational workflow to identify candidate magnetoelectrics and to evaluate the ME tensor in these.
In Sec.~\ref{sec:overview_of_dataset}, we provide an overview of our results and discuss trends in the ME response.
Sec.~\ref{sec:specific_examples_and_subtrends}, discuss specific examples that exemplify different types of response tensors, and in Sec.~\ref{sec:diag_ame}, we highlight examples of compounds having antimagnetoelectric entries in ${\bm \alpha}$.
Finally, in Sec.~\ref{sec:discussion}, we conclude with some general comments and outlook.
The details of the methodology and results are expanded in the Supplemental Material.
We provide the results of DFT calculations for the 124 monolayers as well as the fits to the four independent contributions to ${\bm \alpha}$ for the 3$\times$3 tensor entries.
Although the predicted list of candidates and the evaluated ME response do not contain enough data to identify statistically significant and physical trends, we hope that it may comprise a guide for future studies and stimulate new experimental investigations.
\section{\label{sec:screening_comp_details} Candidate Screening Workflow}

\begin{figure}[t!]\centering
\includegraphics[height=13.75cm]{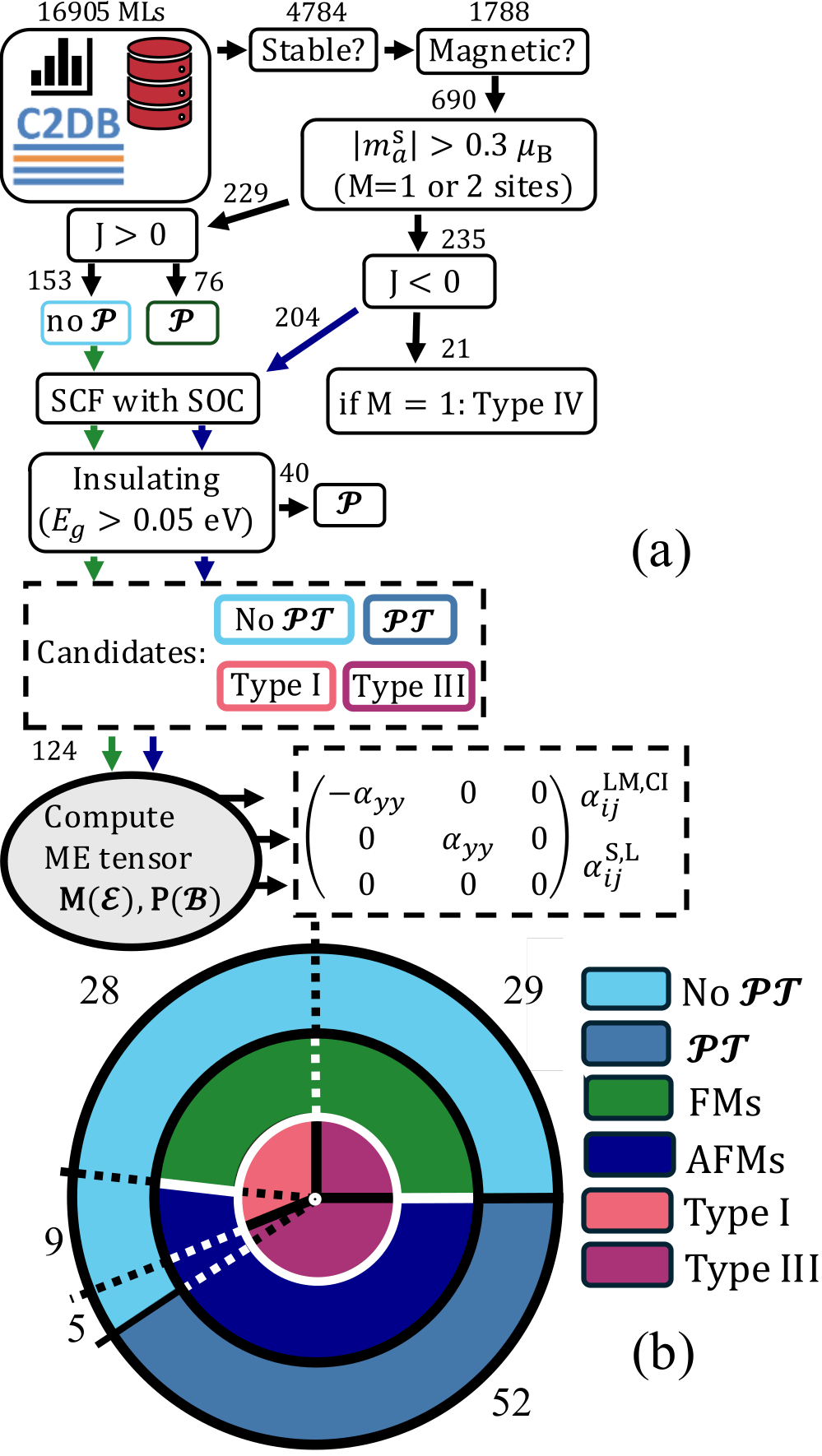}
\caption{\label{fig1} (a) Computational workflow for finding ME monolayers and calculating the ME tensor. (b) Classification of monolayers according to the presence of $\mathcal{PT}$ symmetry, FM or AFM order and type I or type III MSG.
%
% pie chart from http://slid2.fysik.dtu.dk:8888/notebooks/ME_props_v2_C2DB/checkSym_final.ipynb
}
\end{figure}

We perform our DFT calculations using the open source electronic structure package GPAW \cite{Mortensen2024} and the Atomic Simulation Environment (ASE) \cite{Larsen2017}.
All calculations utilize a plane-wave basis set with an energy cut-off of 600 eV and the projector-augmented wave (PAW) method \cite{Blochl1994}. The Brillouin zone was sampled with a 2D uniform $\Gamma$-centered $k$-point grid corresponding to a density of 7 {\AA}$^2$.
The vacuum between periodic images of the monolayers is set to 24 \AA \, in all calculations.
The self-consistent field (SCF) cycles are considered converged when the maximum absolute change in the integrated electronic density is less than $10^{-6}$ $\mathrm{e}^-$/valence electron.
All structures are relaxed to a maximum force of $0.01$ eV /\AA, and for (non-collinear) calculations with self-consistent spin-orbit coupling (SOC), the local density approximation (LDA) exchange correlation functional is used.
Following the workflow outlined in Fig.~\ref{fig1}~(a), we started by selecting the 4784 dynamically stable monolayers (MLs) in C2DB that host one or two magnetic atoms ($M = 1,2$) in their primitive unit cells.
In this regard, the criterion for an atom being classified as magnetic, is a magnetization density that integrates to a value larger (in absolute terms) than $3\mu_\mathrm{B}$ within its PAW sphere.
This initial screening yields 690 magnetic MLs.
Next, we inspect the predicted\cite{Torelli2019} nearest-neighbor isotropic exchange constants $J$ reported using $J \propto E_\mathrm{AFM}^\parallel - E_\mathrm{FM}^\parallel$,
where $E_\mathrm{FM}^\parallel$ and $E_\mathrm{AFM}^\parallel$ are the energies of the collinear FM and AFM configurations.
We find 229 MLs with $J > 0$, which indicates ferromagnetic (FM) ordering. We note that a positive value of the nearest neighbor exchange is not, in itself, sufficient to predict a ferromagnetic ground state, but we will assume so here.
For these materials, we check for the lack of inversion symmetry (no $\mathbfcal{P}$) in the (non-magnetic) space group using \texttt{spglib}\cite{spglibv1} with a distance tolerance of $0.01$ \,\AA.
Since time-reversal symmetry cannot break inversion in ferromagnets we discard all ferromagnets having inversion symmetry in the crystallographic point group.
For the remaining compounds, we perform self-consistent calculations with LDA+SOC and check if the material is insulating.
Our criterion for this is a band gap of $E_g > 0.05$ eV and all materials not satisfying this are discarded.
Finally we find the magnetic space group type using the noncollinear mode of \texttt{spglib}\cite{spglib_mag} with an additional precision tolerance of $0.01$ $\mu_\mathrm{B}$/site.
We point out that $\mathbfcal{T}$ is broken globally in FMs, which usually gives rise to the type I magnetic space group.
However, in some cases, time reversal is coupled to one or more symmetry operations of the group, providing the type III designation.
Next, we turn to the case of $J < 0$ (235 MLs), which typically indicates collinear AFM or non-collinear spin configurations.
In the following, these will be assumed to have a collinear antiferromagnetic ground state.
First, we discard all materials with a single atom in the crystallographic unit cell, since these will belong to the type IV magnetic space groups.
Such materials are characterized by having the symmetry $\mathbfcal{T}\mathbf{t}$ where $\mathbf{t}$ is a translation of a (crystallographic) lattice vector and this will effectively act as time-reversal on $\alpha_{ij}$ and enforce a vanishing linear magnetoelectric response. 
Second, we identify all the remaining materials that lack inversion symmetry in the (non-magnetic) space group, which are guaranteed to exhibit linear magnetoelectric effects. 
However, even compounds having $\mathbfcal{P}$ in the non-magnetic space group may exhibit a linear ME response if inversion symmetry is broken by the magnetic order.
These materials contain the symmetry $\mathbfcal{PT}$ in the magnetic space group and we thus include all such compounds as well (the presence of $\mathbfcal{PT}$  implies the absence of $\mathbfcal{P}$ in the magnetic space group).
This happens, for example, in the case of $\mathrm{Cr}_2\mathrm{O}_3$ discussed in the introduction.
The important distinction between the AFM and FM compounds is that all materials in the C2DB are represented in a FM state and the band gaps always refer to that the FM order.
However, if the proper ground state is AFM, the band gap may change compared to the FM state in C2DB.
For this, we include all materials with $J<0$ from the C2DB and retain the materials fulfilling the symmetry requirements stated above {\it and} exhibit finite band gaps in DFT calculations using the AFM configuration.
With these steps, we find 52 monolayers with $\mathbfcal{PT}$ symmetry and 15 monolayers lacking $\mathbfcal{PT}$ symmetry (broken inversion in the non-magnetic space group).
Finally, the designation of type I or III is assigned by inspecting the corresponding magnetic space group (obtained from \texttt{spglib}) of all the magnetoelectric monolayers.
The final set of monolayers thus allows for three binary classification schemes: FM or AFM, Type I or type III and $\mathbfcal{PT}$ vs no $\mathbfcal{PT}$ symmetry.
We find materials with all combinations of these classes except for the fact that $\mathbfcal{PT}$ is incompatible with ferromagnetism and type I magnetic space groups.

\section{\label{sec:DFT_details} Evaluating the ME response with density functional theory}

Here we will describe the DFT methodology to evaluate ${\bm \alpha}$ focusing on the four aforementioned contributions (${\bm \alpha}^\mathrm{LM,S}$, ${\bm \alpha}^\mathrm{LM,L}$, ${\bm \alpha}^\mathrm{CI,S}$ and ${\bm \alpha}^\mathrm{CI,L}$). 
We provide additional details of our approach in the Supplemental Materials and note that it is also described in detail in our previous work\cite{Mangeri2024}, which also provides a benchmark against reported results on the prototypical ME $\mathrm{Cr}_2\mathrm{O}_3$.
\subsection{\label{sec:thermodynamics} Thermodynamical considerations}
Consider the thermodynamic free energy density ($\mathrm{J}\cdot\mathrm{m}^{-3}$) of a bulk ME (3D) crystal represented as a Taylor series expansion in electric field $\mathbfcal{E}$ ($\mathrm{V}\cdot\mathrm{m}^{-1}$) and magnetic field $\mathbfcal{H}$ ($\mathrm{A}\cdot\mathrm{m}^{-1}$), where first, second, and third-order couplings are stated explicitly\cite{Rivera1994, Weiglhofer2003},
\useshortskip
\begin{align}\label{eq:thermo}
f(\mathbfcal{E}, \mathbfcal{H}) &= ... - P_i^S \mathcal{E}_i - M_i^S \mathcal{H}_i + ...\\ \nonumber
&+ \alpha_{ij} \mathcal{E}_i \mathcal{H}_j \\ \nonumber
&+ \frac{1}{2}\tilde{\beta}_{ijk} \mathcal{E}_i \mathcal{H}_j \mathcal{H}_k  + \frac{1}{2}\beta_{ijk} \mathcal{H}_i \mathcal{E}_j \mathcal{E}_k \\ \nonumber
&+ \frac{1}{2}\tilde{\kappa}_{ijkl} \mathcal{E}_i \mathcal{H}_j \mathcal{H}_k \mathcal{H}_l  + \frac{1}{2}\kappa_{ijkl} \mathcal{H}_i \mathcal{E}_j \mathcal{E}_k \mathcal{E}_l +...\nonumber
\end{align}
where $P_i^S$ and $M_i^S$ are the spontaneous polarization ($\mathrm{C}\cdot\mathrm{m}^{-2}$) and magnetization ($\mathrm{V}\cdot\mathrm{s}^{-1}\cdot\mathrm{m}^{-2}$) and summations over repeated indices are implicit.
It is readily seen that when one takes a variation of $f$ with respect to components of $\mathbfcal{E}$ or $\mathbfcal{H}$, the linear ME coupling ${\bm \alpha}$ becomes the leading order term in $\mathbfcal{E}$ or $\mathbfcal{H}$,
\useshortskip
\begin{align}\label{eq:fit_function_M}
M_r (\mathbfcal{E}) &= -\left(\frac{\partial f}{\partial \mathcal{H}_r}\right)\bigg|_{\mathbfcal{H} = 0} \\ \nonumber
&= M_r^\mathrm{S} - \alpha_{ir} \mathcal{E}_i - \beta_{rjk} \mathcal{E}_j \mathcal{E}_k - \kappa_{rjkl} \mathcal{E}_j \mathcal{E}_k \mathcal{E}_l \nonumber
\end{align}
% \vspace{-1.5em}
%
and
\useshortskip
\begin{align}\label{eq:fit_function_P}
P_r (\mathbfcal{H}) &= -\left(\frac{\partial f}{\partial \mathcal{E}_r}\right)\bigg|_{\mathbfcal{E} = 0} \\ \nonumber
&= P_r^\mathrm{S} - \alpha_{ri} \mathcal{H}_i - \tilde{\beta}_{rjk} \mathcal{H}_j \mathcal{H}_k - \tilde{\kappa}_{rjkl} \mathcal{H}_j \mathcal{H}_k \mathcal{H}_l \nonumber
\end{align}
where we have truncated both expressions to cubic order in the applied field.
In the case of $\mathbfcal{PT}$-symmetric magnets, second-order tensors ${\bm \beta}$ and $\tilde{\bm \beta}$ vanish and the coupling in $\mathbf{M}$ and $\mathbf{P}$ is odd in $\mathbfcal{E}$ and $\mathbfcal{H}$ respectively.
In 2D, $f^\mathrm{2D}$ is given in SI units of $\mathrm{J}\cdot\mathrm{m}^{-2}$ with a 2D $\mathbf{P}$ and $\mathbf{M}$ with units of $\mathrm{C}\cdot\mathrm{m}^{-1}$ and $\mathrm{V}\cdot\mathrm{s}^{-1}\cdot\mathrm{m}^{-1}$.
This changes the units of the various coupling coefficients with ${\bm \alpha}$ carrying SI units of time.
The reader is referred to Refs.~[\citen{Rivera1994, Weiglhofer2003}] for further discussion of units.
We take Eq.~(\ref{eq:fit_function_M}) and (\ref{eq:fit_function_P}), in 2D, as the fitting functions employed in our analysis which is described further in this section.
We note that including higher-order terms provides a more reliable estimate of ${\bm \alpha}$ and is essential for a good fit in some select cases, which we will highlight in Sec.~\ref{sec:specific_examples_and_subtrends}.
A proper treatment of non-linear LM effects requires calculating the higher order Born effective charges, as shown in Ref.~[\citen{Verbeek2023}], but this is beyond the scope of the present work, as we have focused primarily on the linear ME response.

\subsection{\label{sec:lattice_mediated} Lattice-mediated approach}

In the harmonic approximation\cite{Wu2005}, the atomic displacements, $\mathbf{u}$, of a 2D crystal subject to an electric field, $\mathbfcal{E}$, are $u_\beta(\mathbfcal{E}) = \sum_{j\kappa}\left(K_{\beta\kappa}\right)^{-1} Z_{j\kappa}^e \mathcal{E}_j$.
The Greek symbols run over atomic coordinates ($\beta = 1,...,3N_a$) while the Latin indices define the Cartesian reference frame $j = x,y,z$.
The quantities $\mathbf{K}$ and $\mathbf{Z}^e$ are the force constant matrix and the 2D Born effective charges (BECs) which are obtained by finite differences from calculations with displaced atoms (using PBE and a maximum displacement $\pm 0.01$ \AA).
The area of the unit cell is $\mathcal{A}$.
The ground state reference structure was displaced under $\mathbfcal{E}$ in all three Cartesian unit directions $\mathbf{x},\mathbf{y},\mathbf{z}$ (where the special axis $\mathbf{z}||\mathbf{c}$ is always perpendicular to the monolayer plane).
We use a maximum field of $|\mathbfcal{E}| = 5\, \mathrm{mV} \cdot$ \AA${}^{-1}$ that is discretized into 15 linearly separated steps.
Finite differences in the total magnetization per cell area ($\mathbf{M}^\mathrm{S}$) are found from integrating the magnetization density over the entire unit cell and dividing by $\mathcal{A}$ at each $\mathbfcal{E}$.
The ME originates from SOC, which is included self-consistently in all calculations. 
We fit the resulting $\mathbf{M}^\mathrm{S}(\mathbfcal{E})$ to Eq.~(\ref{eq:fit_function_M}) providing the 2D ${\bm \alpha}^\mathrm{LM,S}$ in SI units of seconds.
This results in,
\begin{align}\label{eq:LM_S}
\alpha_{ij}^{\mathrm{LM},\mathrm{S}} \equiv \mu_0 \left(\frac{\partial M_i^\mathrm{S}}{\partial \mathcal{E}_j}\right)\bigg|_{\mathbfcal{B}=0},
\end{align}
where $\mu_0$ is the permeability of vacuum.
Eq.~(\ref{eq:LM_S}) is the first of the contributions (LM,S) to ${\bm \alpha}$.

Next, the orbital moments in the PAW spheres (index $a$) are evaluated with\cite{Ovesen2024, Mortensen2024}, 
\begin{align}\label{eq:orb_mag}
    \mathbf{m}_{a}^\mathrm{L} = \frac{\mu_\mathrm{B}}{N_k} \sum_{\mathbf{k}n} f_{\mathbf{k}n} \left\langle \psi_{\mathbf{k}n}^a\right| \hat{\mathbf{L}}\left| \psi_{\mathbf{k}n}^a \right\rangle,
\end{align}
with $\mu_\mathrm{B}$ the Bohr magneton, $N_k$ the number of $\mathbf{k}$-points, $f_{\mathbf{k} n}$ the band occupancy and $|\psi_{\mathbf{k}n}^a\rangle$ the all-electron wavefunctions represented in terms of a partial wave expansion in the vicinity of the PAW augmentation sphere belonging to atom $a$.
Using the same finite-difference approach as above, we evaluate 
\begin{align}\label{eq:orb_contr}
\alpha_{ij}^{\mathrm{LM},\mathrm{L}} \equiv \mu_0 \left(\frac{\partial M_i^\mathrm{L}}{\partial \mathcal{E}_j}\right)\bigg|_{\mathbfcal{B}=0},
\end{align}
with $\mathbf{M}^\mathrm{L} = \mathcal{A}^{-1} \sum_a \mathbf{m}_a^\mathrm{L}$ to arrive at the second contribution (LM,L) to ${\bm \alpha}$ by fitting the data to Eq.~(\ref{eq:fit_function_M}).
Since we report on a high-throughput study and must balance the precision of the calculations with the need for a large computational load, only two Davidson eigensolver iterations\cite{Davidson1975} are used in the SCF cycle.
However, if the resulting ${\bm \alpha}^\mathrm{LM,(S,L)}$ data appear noisy and do not fit our prescribed cubic polynomial well ($R^2 < 0.8$), we apply five Davidson iterations, which greatly reduce the noise (primarily arising in the integrated magnetization density). 
\subsection{\label{sec:clamped-ion} Clamped-ion approach}
To compute the remaining terms of Eq.~(\ref{eq:total_alpha}), we introduce a Zeeman field $\mathbfcal{B}$ (with $\mathbfcal{E} = 0$) which gives rise to the interaction term $H_z={\bm \sigma}\cdot\mathbfcal{B}$ in the Kohn-Sham Hamiltonian.
We then fix the ionic positions and calculate the total 2D $\mathbf{P}$ using the Berry phase approach \cite{King-Smith1993, Resta1994} while varying the external magnetic field.
The magnetic field is swept from $\pm 0.125$ T in 15 finite intervals in all three Cartesian unit directions $(\mathbf{x},\mathbf{y},\mathbf{z})$.
This amplitude range of $\mathbfcal{B}$ is chosen to avoid possible field-induced spin flop transitions, which gives rise to highly non-linear effects.
For fixed ionic positions, the changes to the electronic degrees of freedom are only coupled to the spin moment through $H_z$, and therefore the applied magnetic field only induces changes to the polarization arising from changes in the spin density.
Hence, 
\begin{align}\label{eq:CI_S}
\alpha_{ij}^{\mathrm{CI},\mathrm{S}} \equiv \mu_0 \left(\frac{\partial P_j}{\partial \mathcal{B}_i}\right)\bigg|_{\mathbfcal{E}=0},
\end{align}
as well as the higher-order couplings can be found by fitting the results to Eq.~(\ref{eq:fit_function_P}).
Since $\mathbf{P}$ is a 2D polarization (dipole moment per area), the resulting SI units are in seconds.
For these calculations, we use five Davidson eigensolver iterations in the SCF cycle to ensure the accuracy required to make $\mathbf{P}$ a smooth function of $\mathcal{B}$.
Finally, we consider the CI contribution due to $\mathbf{M}^\mathrm{L}$ under an applied $\mathbfcal{E}$.
There are a number of ways to compute this, notably using the modern theory of orbital magnetization\cite{Thonhauser2005, Resta2010} and the approach to treat field-polarized Bloch states perturbatively\cite{Souza2002, Malashevich2012}.
We take an alternative path\cite{Mangeri2024}, which is simpler but also likely to be less accurate.
The idea is to apply first-order perturbation theory to Eq.~(\ref{eq:orb_mag}) using an external electrostatic potential $\hat{V} = e \hat{\mathbf{r}}\cdot \mathbfcal{E}$.
The position operator is ill-defined in periodic systems, but interband matrix elements may be expressed in terms of the momentum matrix elements $\langle \psi_{\mathbf{k} m}|\hat{p}_i|\psi_{\mathbf{k} n}\rangle$, which gives
\useshortskip
\begin{align}\label{eq:orb_CIL}
    \alpha_{jk}^\mathrm{CI, L} &\equiv \frac{\mu_0}{\mathcal{A}} \sum\limits_a \left(\frac{\partial m^\mathrm{L}_{a,j}}{\partial \mathcal{E}_k}\right)\bigg|_{\mathbfcal{B}=0} \notag \\
    &= \frac{2i \mu_0 \mu_\mathrm{B}^2}{N_k \mathcal{A}} \sum\limits_a \sum_{\mathbf{k}, n\neq m} (f_{\mathbf{k}n}-f_{\mathbf{k}m}) \notag \\
    &\quad \times \frac{\langle \psi_{\mathbf{k} n}^a|\hat{L}_k| \psi_{\mathbf{k} m}^a\rangle\langle \psi_{\mathbf{k} m}|\hat{p}_j|\psi_{\mathbf{k} n}\rangle}{(\varepsilon_{\mathbf{k}n}-\varepsilon_{\mathbf{k}m})^2}.
\end{align}
The notation $n\neq m$ implies a sum over occupied and unoccupied bands with eigenvalues $\varepsilon_{\mathbf{k}n}$.
%
%The division by $\mathcal{A}$ demonstrates that this quantity is a 2D  tensor (in SI units of seconds).
%
To evaluate this expression, which bears similarity to the ``cross-gap'' term derived in Ref.~[\citen{Essin2010}], we perform a ground state calculation and include additional unoccupied bands (50$\%$ of the number of occupied bands) to ensure convergence of the $n\neq m$ summation.
We have checked that this yields converged numbers for a selected subset of materials.
\subsection{Additional comments}
For each ML, the ME tensor ${\bm \alpha}$ must satisfy
\useshortskip
\begin{align}\label{eq:symmetrize_alpha}
    {\bm \alpha}^\mathrm{sym} = -\mathbfcal{T}_j\,\mathrm{det}(\mathbf{\hat{R}}_j) \mathbf{\hat{R}}_j \, {\bm \alpha}\,\mathbf{\hat{R}}_j^T,
\end{align}
with symmetry operations $\mathbf{\hat{R}}_{j = 1, 2, ...}$ of the MPG and where $\mathbfcal{T}_j$ is the time-reversal operation\cite{spglib_mag, Gallego2019}.
This puts strong constraints on the form of the tensor and the number of inequivalent components.
Using Eq.~(\ref{eq:symmetrize_alpha}), we have evaluated the expected form of the tensor for all monolayers (see Supplemental Materials) with the setting of ${\bm \alpha}^\mathrm{sym}$ given in the C2DB coordinate system (the unique out-of-plane axis is along the $\mathbf{c}$ lattice vector).
Our method for evaluating the ME response may introduce nonzero components $\alpha_{ij}^{(k)}$ which are expected to vanish by symmetry (due to numertical inaccuracies).
However, these components are usually orders of magnitude smaller than those expected to be non-vanishing.
In such cases, we set $\alpha_{ij}$ to zero if $|\alpha_{ij} + \Delta \alpha_{ij}| < \epsilon$.
The quantity $\Delta \alpha_{ij}$ is the sum of the squares of errors obtained from the fit and $\epsilon$ is a tolerance factor (see Supplemental Materials).
With a choice of $\epsilon = 10^{-5}$ attoseconds, along with the convergence criteria mentioned above, we find that in far the majority of MLs, our results reproduce the expected symmetries of ${\bm \alpha}^\mathrm{sym}$ given by Eq.~(\ref{eq:symmetrize_alpha}).
For ${\bm \alpha}^\mathrm{CI,S}$, we sometimes find that the fits are not reliable, which prevents us from making a decisive prediction for that contribution.
Since we do not know the sign of the undetermined $\alpha_{ij}^\mathrm{CI,S}$ relative to other contributions, it is not possible to estimate whether the value for the total ME response is under- or overestimated.
Additional details of the computation of ${\bm \alpha}$ are provided in the Supplemental Materials.

\begin{figure*}[htp!]\centering
\hspace*{-3pt}\includegraphics[height=8.45cm]{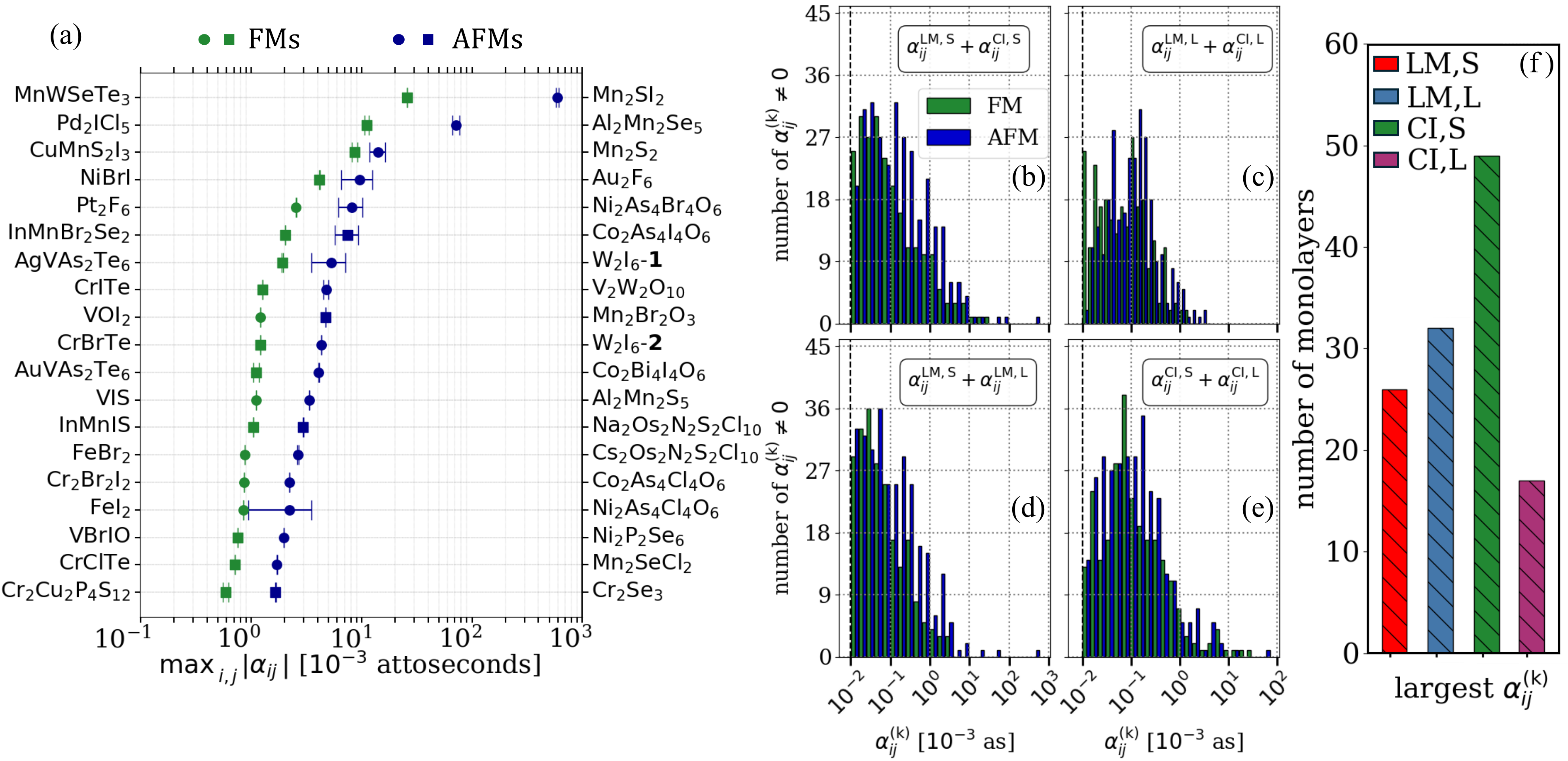}
\caption{\label{fig:fig2} (a) Computed $\alpha_\mathrm{max}$ sorted in ascending order for the largest twenty AFM (to the right) and FM MLs (to the left).
The bolded suffix indicates that the chemical formula has polymorphs (\textbf{1},\textbf{2},...).
The designation of type I and type III magnetic space groups is shown with square and circle markers respectively. The distributions of the spin-driven, orbital-driven, lattice-mediated and clamped-ion decompositions are shown in (b)-(e) respectively.
(f) The data spread of primary contributions to $\alpha_\mathrm{max}$.
}
% http://slid2.fysik.dtu.dk:8888/notebooks/ME_props_v2_C2DB/Untitled.ipynb
\end{figure*}

\section{\label{sec:overview_of_dataset} Summary of results}

Here we comment on some general trends observed in the results.
As we are computing (at most) nine independent tensor components for all $124$ MLs (with four separate contributions each) the data cannot be succinctly presented at once.
However, we provide a Supplemental Material document with all DFT results and the fits for each material/component.
The data is curated in such a way that all four contributions $\alpha_{ij}^{(k)}$ with $(k)$ = LM,S, LM, L, CI,S, CI,L are presented in the same figure and decomposed into the $i,j = x,y,z$ indices.
We also list additional information for each ML, such as the MPG, type I/III magnetic space group, easy axis, band gap $E_g$ (with LDA+SOC), and total ${\bm \alpha}$ as well as the largest $(k)$ contribution and corresponding index $(i,j)$.
Since the primary goal of this work has been to search for materials with a strong linear ME coupling, we compute the largest component of ${\bm \alpha}$ in absolute value for each material as
\useshortskip
\begin{align}\label{eq:max}
\alpha_\mathrm{max} = \mathrm{max}_{i,j} |\alpha_{ij}|.
\end{align}
This evaluation implicitly includes all contributions of Eq.~(\ref{eq:total_alpha}) summed on equal footing (including their relative signs) and in the same unit system (we use $10^{-3}$ attoseconds).
The values of Eq.~(\ref{eq:max}) are displayed in Fig.~\ref{fig:fig2}~(a) for the 20 FM and 20 AFM monolayers having the largest components sorted in ascending order.
The MSG type I or III found from the screening procedure discussed in Sec.~\ref{sec:screening_comp_details} are denoted as squares and circles, respectively.
We also include error bars as estimated from the fitting procedure (see Supplemental Materials for more details).
The strongest response comes from $\mathrm{Mn}_2\mathrm{SI}_2$, which is a $\mathbfcal{PT}$-symmetric AFM.
The largest response of FMs is from $\mathrm{MnWSeTe}_3$.
Both of these MLs have in-plane magnetic order.
We will provide more details of these two compounds (and others) in Sec. \ref{sec:specific_examples_and_subtrends}, which will drive our main point: all contributions to Eq.~(\ref{eq:total_alpha}) may be significant and they should all be evaluated for reliable predictions of linear ME response.
Since a fitting procedure is required to determine ${\bm \alpha}$, we will briefly comment on the fidelity of the results.
$93\%$ of the fits for non-vanishing components have $R^2 > 0.8$ (see Supplemental Materials), which implies that the far majority of results originate from high quality fits.  
A distribution of $R^2$ values is shown in Fig.~S1~(a) and in Fig.~S1~(b) we provide a spread of all relative errors $|\Delta \alpha_{ij} / \alpha_{ij}|$.
The latter of these demonstrates that most compounds have coefficients that are determined with less than $1\%$ error from the fit.
In general, we observe that AFMs tend to exhibit a larger ME response than FMs.
This can be seen in Fig.~\ref{fig:fig2}~(a) where the top 20 AFM MEs exhibit a larger response in $\alpha_\mathrm{max}$.
We can characterize the distribution of materials that are spin or orbital driven by computing the sums $\alpha_{ij}^\mathrm{LM,S} + \alpha_{ij}^\mathrm{CI,S}$ or $\alpha_{ij}^\mathrm{LM,L} + \alpha_{ij}^\mathrm{CI,L}$, respectively, as shown in Fig.~\ref{fig:fig2}~(b) and (c).
Similarly, we decompose the lattice-mediated and clamped-ion effects by providing $\alpha_{ij}^\mathrm{LM,S} + \alpha_{ij}^\mathrm{LM,L}$ and $\alpha_{ij}^\mathrm{CI,S} + \alpha_{ij}^\mathrm{CI,L}$, respectively, in (d) and (e).
Here, one sees that, in general, all of these contributions will be a significant factor in the total $\alpha_{ij}$ and the orbital driven MLs typically have larger responses.
We can also analyze the results in terms of the largest of the four contributions for all materials.
This is shown in Fig. \ref{fig:fig2}(f), and it is observed that 21$\%$ of the MLs have LM,S as the principal contribution while $26\%, 40\%, 13\%$ have LM,L CI,S and CI,L as the dominant contributions, respectively.
The information in (b)-(f) quantitatively demonstrates that if one wants to study the ME response from the level of DFT, one needs to include all contributions. 
We emphasize this point in Sec.~\ref{sec:specific_examples_and_subtrends} by specific examples.
We finally comment on the physical correlations in the results.
In the LM mode, the applied $\mathbfcal{E}$ results in nuclear coordinate displacements and the magnitude of these are expected to be rooted in the softness of the optical phonons and the strength of the Born effective charges.
Therefore, one might assume that the strength of LM response should correlate with the lowest optical phonon energy at $\Gamma$ as well as the $Z_{j\kappa}^e$ entries.
However, we did not manage to identify such correlation.
We also investigated if $\alpha_\mathrm{max}$ is correlated with the SOC energy, the band gap, the calculated nearest neighbor exchange $J$, the number of nearest neighbors $N$ as well as the nearest neighbor distances.
In all cases, we were unable to identify any clear correlation.
We leave such questions for later work and emphasize that a larger dataset would be beneficial for uncovering physical trends that underpin the ME response.
Furthermore, we should mention that there is no influence of the classification criteria [MSG type I/III or the presence (or lack) of $\mathbfcal{PT}$ symmetry] on the results.

\begin{figure}[htp!]\centering
\includegraphics[height=6.2cm]{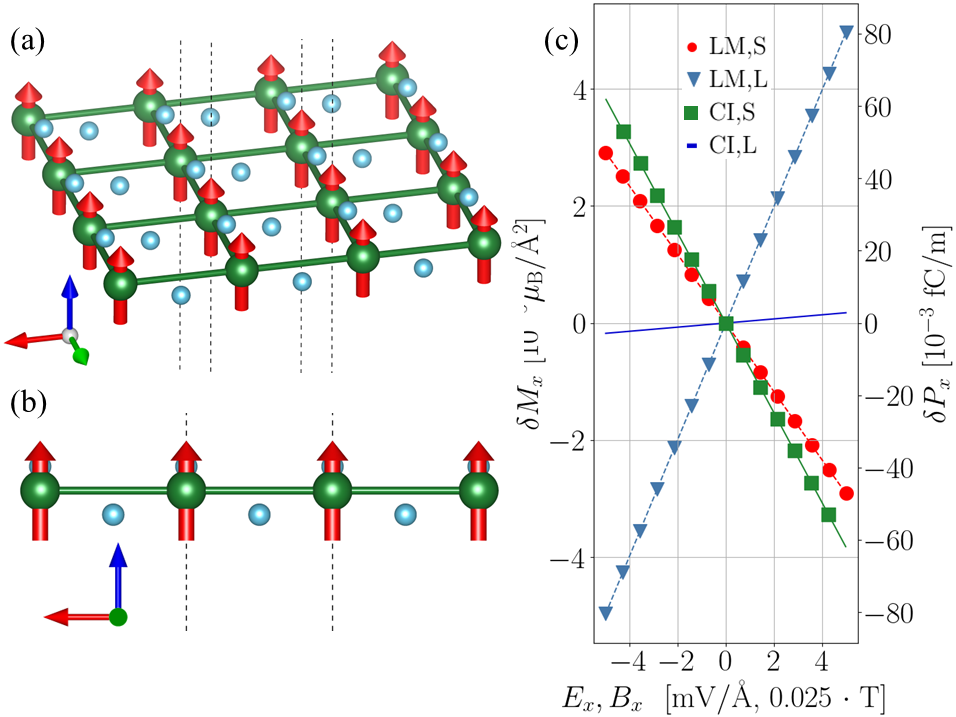}
\caption{\label{fig:tao2} A perspective and side view of the FM $\mathrm{TaO}_2$ monolayer in (a) and (b) respectively.
Tantalum and oxygen ions are in green and light blue correspondingly.
In (c), the ME response data ($\alpha_{xx} = - \alpha_{yy}$) for LM,S (red circles), LM,L (light blue triangles), CI,S (green squares) with fits presented as lines along with our estimate of CI,L (blue line).}
\end{figure}

\section{\label{sec:specific_examples_and_subtrends} Specific Examples}

Here we exemplify our results by analyzing four monolayers in detail.
We first consider the FM compound $\mathrm{TaO}_2$, which has the parent paramagnetic space group $P\bar{4}m2$ and an out-of-plane easy axis.
Inversion is broken (no $\mathbfcal{P}$) by alternating oxygens positioned above and below the square Ta-Ta bond network.
The structure and magnetic order are visualized\cite{Momma2011} in Fig.~\ref{fig:tao2}~(a) and (b).
The magnetic point group of this configuration is $4'/m'm'm$ which dictates that the tensor has the structure
\begin{align}\label{eq:alpha_mpt4mmm}
\renewcommand{\arraystretch}{0.5}
{\bm \alpha}^\mathrm{sym}(4'/m'm'm) =  \begin{pmatrix}
\alpha_{xx} & 0 & 0\\
  &  & \\
0 & -\alpha_{xx} & 0\\
  &  & \\
0 & 0 & 0 \\
\end{pmatrix},
\end{align}
with only one independent coefficient.
Following the procedures outlined in Sec.~\ref{sec:DFT_details}, the $\{\mathbfcal{E},\mathbfcal{B}\}$ fields are applied along the Cartesian $\mathbf{x}$ direction and $\delta M_x^\mathrm{S}, \delta M_x^\mathrm{L}$ and $\delta P_x$ (deviations from the ground state) are plotted in Fig.~\ref{fig:tao2}~(c) as red circles, blue triangles, and green squares respectively.
The resulting fits are shown as lines - and the expression computed from Eq.~(\ref{eq:orb_CIL}) is overlayed on the figure as a solid blue line in units ($\mu_\mathrm{B}/$\AA${}^2$) of the left-hand vertical axis.
From our calculations, we find the total $\alpha_{xx} = -0.11$ whereas $\alpha_{yy} = +0.11$ in units of $10^{-3}$ attoseconds.
This agrees very well with the expected ${\bm \alpha}^\mathrm{sym}$ according to Eq.~(\ref{eq:alpha_mpt4mmm}).
The largest contribution is from $\alpha_{xx}^\mathrm{LM,L} = 1.16\times$ $10^{-3}$ attoseconds. 
However, $\alpha_{xx}^\mathrm{LM,S}+\alpha_{xx}^\mathrm{CI,S} = -0.68\times 10^{-3}+(-0.62\times 10^{-3}) = -1.30\times 10^{-3}$ attoseconds.
The remaining term, $\alpha_{xx}^\mathrm{CI,L} = 0.04 \times 10^{-3}$ attoseconds, is much weaker.
Therefore, the sum of all terms from Eq.~(\ref{eq:total_alpha}) sum to a negative value.
If only $\alpha_{xx}^\mathrm{LM,S}$ was considered, the predicted value of ${\bm \alpha}$ would be overestimated by a factor of six and have the wrong sign.
In this system, it is evident that different contributions to ${\bm \alpha}$ effectively cancel out.

\begin{figure}[htp!]\centering
\includegraphics[height=6.0cm]{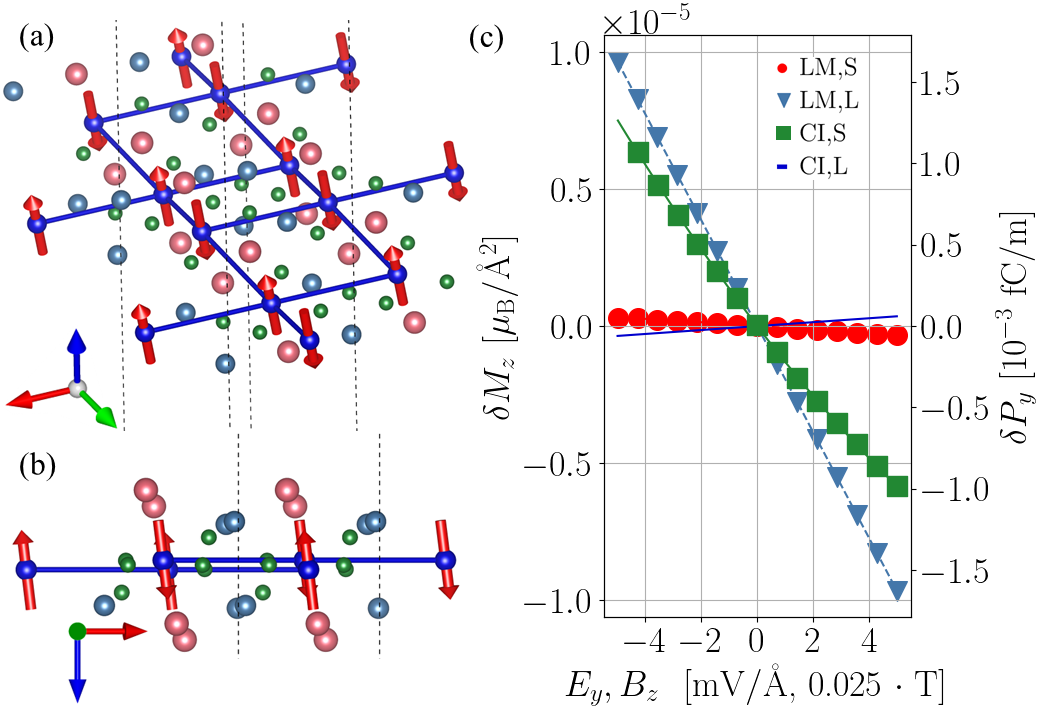}
\caption{\label{fig:ni2br4sb4O6} A persepective (a) and side (b) view of AFM monolayer $\mathrm{Ni}_2\mathrm{Br}_4\mathrm{Sb}_4\mathrm{O}_6$ and the (c) ME response ($\alpha_{zy}$) for LM,S (red circles), LM,L (light blue triangles), CI,S (green squares) with fits presented as lines along with our estimate of CI,L (blue line).}
\end{figure}

Next, we consider the AFM ML $\mathrm{Ni}_2\mathrm{Br}_4\mathrm{Sb}_4\mathrm{O}_6$ with paramagnetic parent symmetry of $P\bar{1}$ and magnetic point group of $\bar{1}'$.
The only symmetry is thus $\mathbfcal{PT}$ and the easy axis is largely out-of-plane with a small canting along the $\mathbf{x}$ direction.
The perspective and side view of the structure with magnetic moments is visualized in Fig.~\ref{fig:ni2br4sb4O6}~(a) and (b) respectively.
The low symmetry allows for nine independent coefficients in the response tensor:
\begin{align}\label{eq:alpha_m1}
\renewcommand{\arraystretch}{0.5}
{\bm \alpha}^\mathrm{sym}(\bar{1}') =  \begin{pmatrix}
\alpha_{xx} & \alpha_{yx} & \alpha_{zx}\\
  &  & \\
\alpha_{xy} & \alpha_{yy} & \alpha_{zy}\\
  &  & \\
\alpha_{xz} & \alpha_{yz} & \alpha_{zz} \\
\end{pmatrix}.
\end{align}
With the same prescription as the case above and with fields $\mathbfcal{E}||\mathbf{y}$ and $\mathbfcal{B}||\mathbf{z}$, we plot the data for out-of-plane $M_z^\mathrm{S}, M_z^\mathrm{L}$ and $P_y$ in (c).
In this coefficient, the total ${\bm \alpha}$ is predominantly driven by LM,L ($-0.23\times10^{-3}$ attoseconds) whereas all other coefficients are on the order of $0.01\times 10^{-3}$ attoseconds.
If the orbital magnetization was neglected, the response would have been predicted to be nearly zero and this exemplifies a case where the primary origin of the ME effect is orbital magnetization.

\begin{figure}[htp!]\centering
\includegraphics[height=5.8cm]{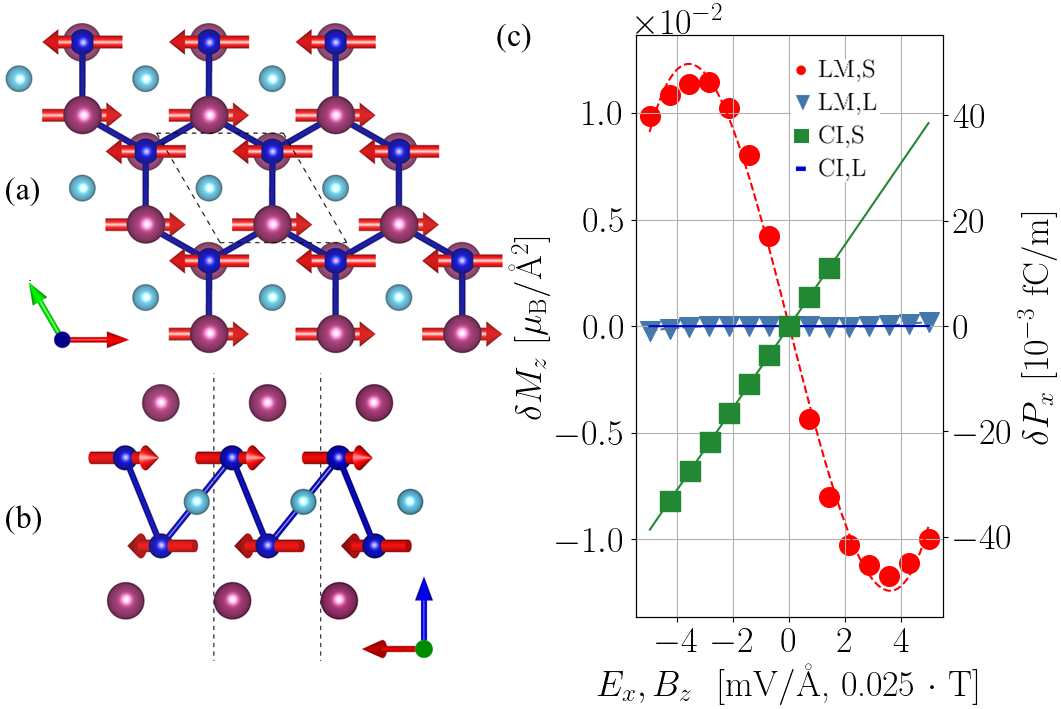}
\caption{\label{fig:Mn2SI2} Top (a) and side view (b) of magnetic order and structure in AFM $\mathrm{Mn}_2\mathrm{SI}_2$.(c) ME response $(\alpha_{zx})$ for LM,S (red circles), LM,L (light blue triangles), CI,S (green squares) and CI,L (blue line).}
\end{figure}

We now consider the largest ME response in the results.
This is observed in the AFM ML $\mathrm{Mn}_2\mathrm{SI}_2$,  a $\mathbfcal{PT}$-symmetric magnet with (non-magnetic) space group $P\bar{3}m1$.
The magnetic moments order in-plane along $\mathbf{x})$, which yields the magnetic point group $\bar{1}'$.
Thus, there are again nine independent non-vanishing coefficients.
We provide a visualization of the honeycomb structure and in-plane magnetic order in Fig. \ref{fig:Mn2SI2} (a) and (b).

The out-of-plane components $M_z^\mathrm{S}, M_z^\mathrm{L}$ and the in-plane components $P_x$ as a function of $\mathbfcal{E}||\mathbf{x}$ and $\mathbfcal{B}||\mathbf{z}$, respectively, are shown in Fig.~\ref{fig:Mn2SI2}~(c) 
The resulting fits to Eq.~(\ref{eq:fit_function_M}) and (\ref{eq:fit_function_P}) are shown as lines.
Similar to the two previous examples, we find that $\alpha_{zx}^\mathrm{CI,L}$ is rather small ($0.08\times 10^{-3}$ attoseconds) and do not provide an important contribution to the ME response.
All remaining contributions are all rather large though with $\alpha_{zx}^\mathrm{LM,L}$ is $-2.7\times 10^{-3}$ attoseconds, which is comparable to clamped-ion spin contribution of $\alpha_{zx}^\mathrm{CI,S} = 1.27\times 10^{-3}$ attoseconds.
However, both of these contributions are insignificant compared to the lattice mediated spin, which yields $\alpha_{zx}^\mathrm{LM,S}-599.0\times 10^{-3}$ attoseconds - the highest value found in all of the considered materials.
We point out that a cubic dependence of $M_z^\mathrm{S}$ on $\mathcal{E}_z$ is evident from Fig. \ref{fig:Mn2SI2} and the fitted {\it linear} contribution is greatly improved by including the higher order terms of Eq.~(\ref{eq:fit_function_M}), ($R^2 = 0.994$ with an uncertainty $\Delta\alpha_{zx}^\mathrm{LM,S} = \pm 20\times 10^{-3}$ attoseconds).
%
%Such a large value is quite surprising, as all other components of $\alpha_{ij}$ are one or two orders of magnitude smaller (see Fig.~S90 in the Supplemental Materials).
%
This large value is likely to originate from the large spin density (10 $\mu_\mathrm{B}$ per unit cell) as well as a weak magnetic anisotropy that makes the moments prone to cant into the $z$-direction given a small distortion.
Since this is the strongest ME response in the dataset, it is instructive to compare it to measured values of other bulk compounds in the literature.
However, this requires converting the 2D ME to 3D bulk values, which can be accomplished by assuming a van der Waals bonded compound with a typical distance between layers.
The thickness of the layer is $h_\mathrm{ML} = 6.33\,$ \AA  (computed by taking the differences of the maximum and minimum atomic position along the $\mathbf{c}$-axis) 
and we then add approximately $h_\mathrm{vDW} = 4$ \,\AA, which is a reasonable van der Waals gap.
Using $\alpha_{zx}^\mathrm{3D} = \alpha_{zx} / (h_\mathrm{ML} + h_\mathrm{vDW})$, we find $\alpha_{zx}^\mathrm{3D} \approx $ 580 $\mathrm{ps}\cdot\mathrm{m}^{-1}$.
%
% Can add uncertainty...
%
This value is two orders of magnitude larger than the transverse response in the prototypical ME $\mathrm{Cr}_2\mathrm{O}_3$ ($\alpha_\perp \simeq 1.2$ ps$\cdot\mathrm{m}^{-1}$), but is comparable to the largest ME response measured in $\mathrm{TbPO}_4$ ($\alpha \simeq 280-730$ ps$\cdot\mathrm{m}^{-1}$)\cite{Rivera2009, Grams2024}.

\begin{figure}[htp!]\centering
\includegraphics[height=6.00cm]{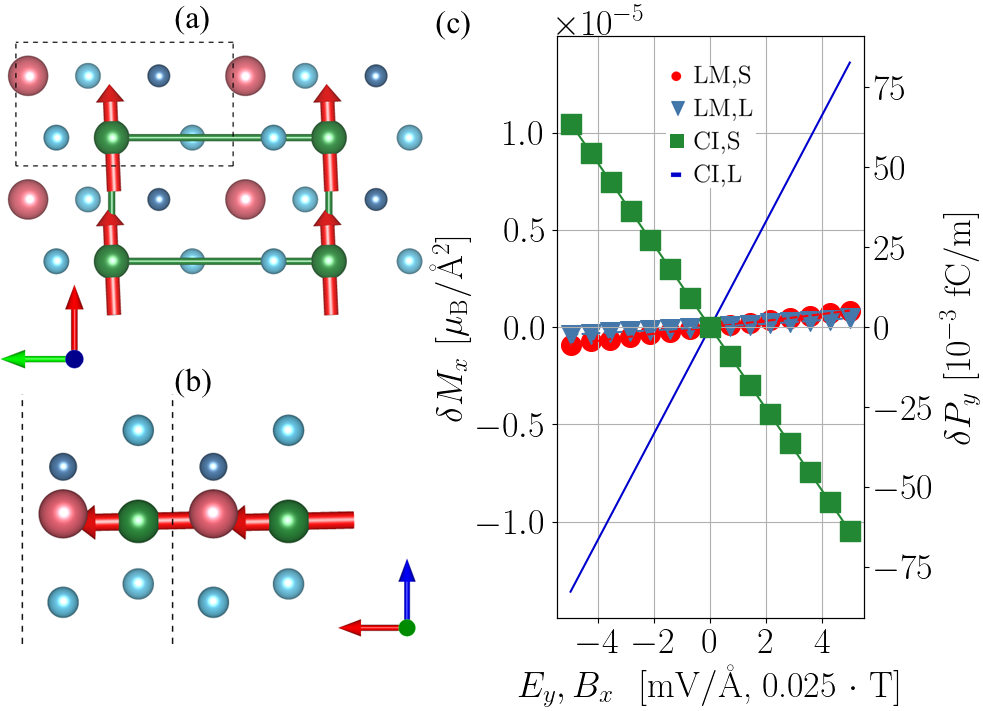}
\caption{\label{fig:mnsewte3} Top (a) and side-view (b) for FM $\mathrm{MnSeWTe}_3$. (c) ME response ($\alpha_{xy}$) for LM,S (red circles), LM,L (light blue triangles), CI,S (green squares) and CI,L (blue line).}
\end{figure}

As our final example, we consider the FM case of ML $\mathrm{MnWSeTe}_3$ which crystallizes into the tetragonal non-magnetic space group $Pm11$.
The magnetic easy axis is along the short in-plane direction ($\mathbf{a}||\mathbf{x}$), which reduces the symmetry, and the the MPG becomes $1$.
We present the spin and structure configuration in (a) and (b) of Fig.~\ref{fig:mnsewte3} and expect nine independent components of the linear ME tensor similar to the two previous examples.
In (c), $M_x^\mathrm{S}, M_x^\mathrm{L}$ and $P_y$ are shown as a function of $\mathbfcal{E}||\mathbf{y}$ and $\mathbfcal{B}||\mathbf{x}$ respectively.
The fits yields values of +0.01 and +0.02 in units of $10^{-3}$ attoseconds for $\alpha_{xy}^\mathrm{LM,S}$ and $\alpha_{xy}^\mathrm{LM,L}$ respectively.
These contributions are negligible compared to the clamped-ion contributions $\alpha_{xy}^\mathrm{CI,S}=-26.0\times 10^{-3}$ and $\alpha_{xy}^\mathrm{CI,L}=0.32\times 10^{-3}$ attoseconds.
This component (which is the strongest of all of the FMs) is thus completely dominated by the clamped-ion spin term.
Assuming a van der Waals gap of 4 \,\AA, we find $\alpha_{xy}^\mathrm{3D} \approx -33$ $\mathrm{ps}\cdot\mathrm{m}^{-1}$.
This material exemplifies the most common situation in our data, that CI,S (electronic) effects constitute the primary response in $\alpha_\mathrm{max}$ [see Fig.~\ref{fig:fig2}~(f)].
Several MLs in our dataset have been assigned ICSD or COD identifiers implying that they are experimentally known as bulk van der Waals bonded compounds.
This suggests that they can be (or already have been) exfoliated to the 2D limit\cite{Carretta2002,  Hugonin2008, Baithi2023, Peng2020, Lai2019, Selter2023, Matsuoka2024},  and such monolayers may thus be of particular interest.
In Table~\ref{tab:table1}, we summarize these monolayers for easy reference.
In this regard, it should be noted that the sign of ${\bm \alpha}$ is determined by the orientation of the magnetic domain\cite{Malashevich2012, Bousquet2024}.
Hence, in samples with an equal population of opposite sign domains, the effective ${\bm \alpha}$ will be zero, making experimental measurements of ${\bm \alpha}$ challenging.

\begin{table}[htp!]
\begin{ruledtabular}
\begin{tabular}{c c c c c}
    Formula (MPG)   &  \hspace*{-5pt}${\bm \alpha}^\mathrm{sym}$  & \hspace*{-5pt}${\bm \alpha}$ \\
   % & &    & \\
    \hline
 \hspace*{-10pt} $\mathrm{Mo}_2\mathrm{V}_2\mathrm{O}_{10}$ ($4/m'$)&  \hspace*{-10pt}$\begin{pmatrix}
      \alpha_{xx} & \alpha_{yx} & 0 \\
      -\alpha_{yx} & \alpha_{yy} & 0 \\
      0 & 0 & \alpha_{zz} 
      \end{pmatrix}$ &  \hspace*{-8pt}$\begin{pmatrix}
      0.27 & -0.03 & 0 \\
      0.03 &  0.27& 0 \\
      0 & 0 & 0.07
      \end{pmatrix}$ \\
    COD 1535988   & &    & \\
       & &    & \\
 \hspace*{-10pt} $\mathrm{AgVP}_2\mathrm{Se}_6$ ($2'$) &  \hspace*{-10pt}$\begin{pmatrix}
     -\alpha_{yy} & \alpha_{yx} & \alpha_{yz} \\
      -\alpha_{yx} & \alpha_{yy} & -\alpha_{yz} \\
      -\alpha_{yz} & \alpha_{yz} & 0 
      \end{pmatrix}$ &  \hspace*{-8pt}$\begin{pmatrix}
      -0.04 & 0.23 & 0.02 \\
      -0.29 &  0.06& -0.02 \\
      -0.01 & 0.05 & 0
      \end{pmatrix}$ \\
    COD 1509506   & &    & \\
       & &    & \\
 \hspace*{-10pt} $\mathrm{Cr}_2\mathrm{P}_2\mathrm{Se}_6$ ($m$)&  \hspace*{-20pt}$\begin{pmatrix}
     0 & \alpha_{yx} & 0 \\
    \alpha_{xy} & 0 & \alpha_{zy} \\
     0 & \alpha_{yz} & 0 
      \end{pmatrix}$ &  \hspace*{-15pt}$\begin{pmatrix}
      0 & -0.03 & 0 \\
      -0.23 &  0& -0.02 \\
      0 & -0.07 & 0.0
      \end{pmatrix}$ \\
    ICSD 626521  & &    & \\
    %   & &    & \\
 \hspace*{-10pt} $\mathrm{Co}_2\mathrm{Br}_4\mathrm{Sb}_4\mathrm{O}_6$ ($-1'$) &   \hspace*{-10pt}$\begin{pmatrix}
     \alpha_{xx} & \alpha_{yx} & \alpha_{zx} \\
    \alpha_{xy} & \alpha_{yy} & \alpha_{zy} \\
     \alpha_{xx} & \alpha_{yz} & \alpha_{zz} 
      \end{pmatrix}$ &  \hspace*{-8pt}$\begin{pmatrix}
      -0.06 & 0.05 & 0 \\
      0.01 &  -0.31 & 0 \\
      -0.15 & -0.01 & 0.04
      \end{pmatrix}$ \\
    ICSD 418858  & &    & \\
    %   & &    & \\ %this one has some rounding issues in final plot. Fix!
 \hspace*{-10pt} $\mathrm{NiCl}_2\mathrm{N}_2\mathrm{H}_4\mathrm{C}_6$ ($m'm'2$) &   \hspace*{-10pt}$\begin{pmatrix}
     \alpha_{xx} & 0 & 0 \\
    0 & \alpha_{yy} & 0 \\
    0 & 0 & \alpha_{zz} 
      \end{pmatrix}$ &  \hspace*{-5pt}$\begin{pmatrix}
      0.03 & 0 & 0 \\
      0 &  -0.01 & 0 \\
      0 & 0 & 0
      \end{pmatrix}$ \\
    COD 7227895  & &    & \\
     %  & &    & \\
 \hspace*{-10pt} $\mathrm{Ni}_2\mathrm{P}_2\mathrm{Se}_6$ ($-3'm$) &   \hspace*{-10pt}$\begin{pmatrix}
     0 & \alpha_{yx} & 0 \\
    -\alpha_{yx} & 0 & 0 \\
    0 & 0 & 0
      \end{pmatrix}$ &  \hspace*{-5pt}$\begin{pmatrix}
      0 & -2.00 & 0.01 \\
      1.94 & 0 & 0 \\
      0 & 0 & 0
      \end{pmatrix}$ \\
    Ref.~[\citen{Matsuoka2024}]  & &    & \\ 
    %
%also can add
     %V2Se2H4O10
     %ICSD id of parent bulk structure 	69994
\end{tabular}
\end{ruledtabular}
\caption{\label{tab:table1}%
Exfoliable materials with their COD, ICSD or reference article.
Expected ${\bm \alpha}$ by the MPG symmetry (also listed) as well as our computed total value.
}
\end{table}

\section{\label{sec:diag_ame} Antimagnetoelectric tensor entries}

\begin{figure}[htp!]\centering
\includegraphics[height=9.50cm]{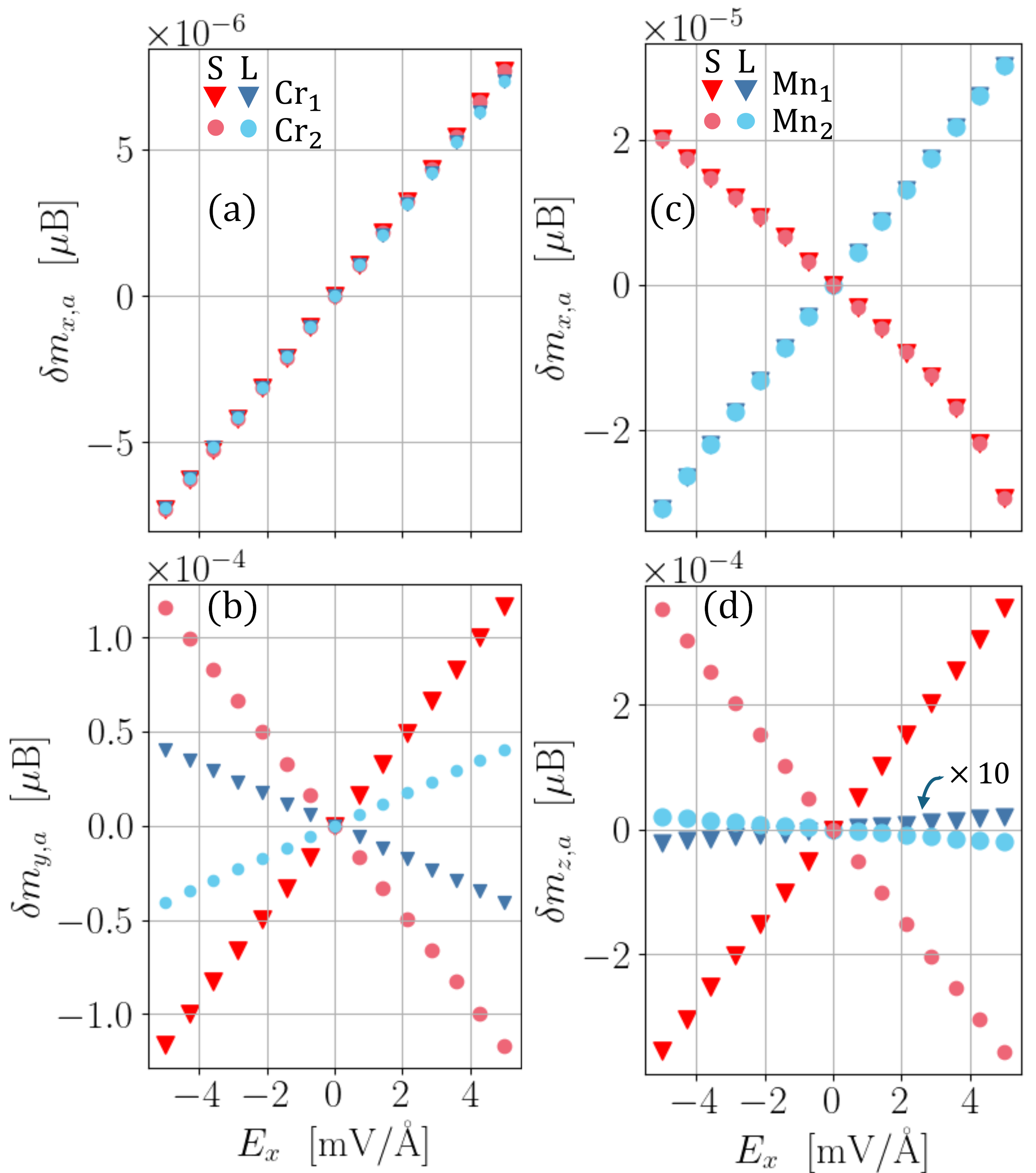}
\caption{\label{fig:ame} Local spin (red triangles, pink circles) and orbital (blue triangles, light blue circles) of the two magnetic atoms respectively for $\delta m_{x,a}$ (a) and $\delta m_{y,a}$ (b) of $\mathrm{Cr}_2\mathrm{As}_2\mathrm{S}_6$. In (c) and (d) are $\delta m_{x,a}$ and $\delta m_{z,a}$ respectively for $\mathrm{Mn}_2\mathrm{S}_2\mathrm{H}_4\mathrm{O}_8$. In (d), we amplify the orbital moment by a factor of 10 for visualization.}
%\textbf{J: low-res, will fix}
%plot and analysis is in http://slid2.fysik.dtu.dk:8888/notebooks/ME_props_v2_C2DB/find_anti_MEs.ipynb
\end{figure}

One phenomenon that we find in our data relates to the concept of anti-magnetoelectricity \cite{Verbeek2023}.
The anti-magnetoelectric (anti-ME) effect is a hidden response present in MEs where the local site symmetry of the magnetic site preserves a nonzero \emph{local} ME excitation of a given ${\bm \alpha}$ component.
However, the total response of the crystal, for a given entry $\alpha_{ij}$, vanishes since two sites may have equal and opposite induced moments.
To illustrate this, we first consider the case of $\mathrm{Cr}_2\mathrm{O}_3$ highlighted in Ref.~[\citen{Verbeek2023}].
The form of ${\bm \alpha}$ due to the MPG $3m'$ (in the appropriate $\mathbf{S}||\hat{\mathbf{z}}$ setting) is
\begin{align}\label{eq:alpha_cro}
\renewcommand{\arraystretch}{0.5}
{\bm \alpha}(3m') =  \begin{pmatrix}
\alpha_{xx} & 0 & 0\\
  &  & \\
0 & \alpha_{xx} & 0\\
  &  & \\
0 & 0 & \alpha_{zz} \\
\end{pmatrix}.
\end{align}
Locally, the two nonequivalent Cr atoms in the primitive magnetic cell exhibit off-diagonal linear ME responses, as demonstrated by a multipole expansion and DFT calculations\cite{VerbeekThesis2024}, which yields
\begin{align}\label{eq:alpha_cro_1}
\renewcommand{\arraystretch}{0.5}
{\bm \alpha}^{\mathrm{Cr}1} =  \begin{pmatrix}
\alpha_{xx}^{\mathrm{Cr}} & \alpha_{yx}^{\mathrm{Cr}} & 0\\
  &  & \\
-\alpha_{yx}^{\mathrm{Cr}} & \alpha_{xx}^{\mathrm{Cr}} & 0\\
  &  & \\
0 & 0 & \alpha_{zz}^{\mathrm{Cr}1} \\
\end{pmatrix}
\end{align}
and
\begin{align}\label{eq:alpha_cro_2}
\renewcommand{\arraystretch}{0.5}
{\bm \alpha}^{\mathrm{Cr}2} =  \begin{pmatrix}
\alpha_{xx}^{\mathrm{Cr}} & -\alpha_{yx}^{\mathrm{Cr}} & 0\\
  &  & \\
\alpha_{yx}^{\mathrm{Cr}} & \alpha_{xx}^{\mathrm{Cr}} & 0\\
  &  & \\
0 & 0 & \alpha_{zz}^{\mathrm{Cr}} \\
\end{pmatrix}.
\end{align}
Here, off-diagonal components of ${\bm \alpha} = {\bm \alpha}^{\mathrm{Cr}1}+{\bm \alpha}^{\mathrm{Cr}2}$ exhibit an \emph{antiferroic} character while the diagonal entries preserve the linear ME effect.
For this analysis, the oxygen atoms are neglected as they also display antiferroic response albeit with much weaker moments.
A similar picture emerges for hematite $\alpha$-$\mathrm{Fe}_2\mathrm{O}_3$, which has the same inversion-symmetric corundrum structure as $\mathrm{Cr}_2\mathrm{O}_3$.
The magnetic ordering, however, is different, yielding a MPG of $\bar{3}m'$, which does not contain $\mathbfcal{PT}$ and all coefficients of the total ${\bm \alpha}$ vanish.
In Ref.~[\citen{Verbeek2023}], it was shown that Fe atom-resolved components of ${\bm \alpha}$ in hematite also display antiferroic responses of equal magnitude and opposite sign demonstrating that this material is an anti-ME.
This is because the inherent inversion symmetry involves exchange of two Fe atoms and the inversion symmetry allows for finite response at a particular site.
We investigated our results to identify whether any of the vanishing LM coefficients are of anti-ME character.
For this, we computed the local magnetic (spin) moments, $\mathbf{m}^\mathrm{S}_a$, found by integrating the magnetization density in the PAW spheres.
Additionally, we use Eq.~(\ref{eq:orb_mag}) to find the local orbital magnetic moments in the atom-centered approximation ($\mathbf{m}^\mathrm{L}_a$) to probe if the orbital part can also be anti-ME.
As an example, we consider the FM ML $\mathrm{As}_2\mathrm{Cr}_2\mathrm{S}_6$ which has out-of-plane easy axis and the MPG $3m'$ giving a diagonal tensor of the form~(\ref{eq:alpha_cro}).
In Fig.~\ref{fig:ame}~(a), we show $\delta m_{x,a}^\mathrm{S}$ (red triangles, pink circles) and $\delta m_{x,a}^\mathrm{L}$ (blue triangles, light blue circles) data for the two Cr atoms, respectively.
We see that the induced moments under $\mathcal{E}_x$ all have the same sign.
%
%urprisingly, they all have the same magnitude as well showing that $\alpha_{xx}^\mathrm{LM,S}\approx\alpha_{xx}^\mathrm{LM,L}$.
%
A different picture emerges when looking at (b), which shows $\delta m_{y,a}^\mathrm{S}$ and $\delta m_{y,a}^\mathrm{L}$ under $\mathcal{E}_x$.
Here, the moments of each site are of equal magnitude and opposite in sign.
This is exactly the same case as $\mathrm{Cr}_2\mathrm{O}_3$, in that the local $\mathrm{Cr}$ atoms follow Eq.~(\ref{eq:alpha_cro_1}) and (\ref{eq:alpha_cro_2}).
%
%In this analysis, we neglect contributions from the ligands, which are small.
%
Therefore, with this data, we directly observe the anti-ME effect in the $\alpha_{yx}$ component, but not in $\alpha_{xx}$.
Additionally, the orbital response also has anti-ME character as shown in the blue curves of (b) which is another observation not commented on previously in the literature.
Another remarkable result is that the magnitude of the induced site moments (both spin and orbital) in the anti-ME $\alpha_{yx}^\mathrm{LM}$ component are two orders of magnitude larger than those in the nonzero $\alpha_{xx}^\mathrm{LM}$ for both spin and orbital contributions.

As a second example, we consider the AFM ML $\mathrm{Mn}_2\mathrm{S}_2\mathrm{H}_4\mathrm{O}_8$ that is without $\mathbfcal{PT}$.
In this compound, the moments order along the $\hat{\mathbf{z}}$ axis.
The computed MPG is $m'$ yielding,
\begin{align}\label{eq:alpha_32p}
\renewcommand{\arraystretch}{0.5}
{\bm \alpha}^\mathrm{sym}(m') =  \begin{pmatrix}
\alpha_{xx} & \alpha_{yx} & 0\\
  &  & \\
\alpha_{xy} & \alpha_{yy} & 0\\
  &  & \\
0 & 0 & \alpha_{zz} \\
\end{pmatrix}.
\end{align}
In Fig.~\ref{fig:ame} (c), we display the local orbital moments of the $\mathrm{Mn}$ sites relating to the $\alpha_{xx}^\mathrm{LM}$ component in the same marking scheme as (a) and (b).
Here, both moments of the $\mathrm{Mn}$ sites are the same sign and magnitude for spin and orbital contributions respectively.
However, since the spin response does not completely cancel the orbital one with opposite sign, there is a net $\alpha_{xx}^\mathrm{LM}$.
In (d), we show the anti-ME response of $\delta m_{z,a}^\mathrm{S}$ and $\delta m_{z,a}^\mathrm{L}$ where each $\mathrm{Mn}$ site has equal magnitude with opposite sign.
We amplify $\delta m_{z,a}^\mathrm{L}$ by a factor of 10 for visualization - showing that it also has anti-ME character but with much smaller amplitude relative to the spin component.
Thus, both LM,S and LM,L contributions vanish in $\alpha_{zx}$ as a consequence of the anti-ME coupling.
The clamped-ion terms are not found through local approximations and therefore we cannot find an analog to this effect, although it would be an interesting phenomenon to unravel.
These two examples are similar to $\mathrm{Cr}_2\mathrm{O}_3$ (anti-ME entries in a material with global symmetry that provides a non-zero ME effect).
We do not conduct a thorough investigation of the anti-ME entries in this work, but note that it is, in general, \emph{not} present in all entries that are expected to be zero.
Instead, it only appears in distinct entries of ${\bm \alpha}$.
Our screening procedure described in Sec.~\ref{sec:screening_comp_details} neglects materials that would have anti-ME effects in all coefficients, as is the case of $\alpha$-$\mathrm{Fe}_2\mathrm{O}_3$.
A more comprehensive study of anti-ME entries is beyond the scope of this work and we leave this for future endeavors.

It should finally be noted that the site resolved response is not completely well defined for the simple reason that the local magnetic moments are not well defined.
This renders the anti-ME response somewhat ill-defined at the quantitative level although site symmetries may be applied to strictly predict the presence or absence of the effect in a particular component. Moreover, the magnetization density is typically rather strongly localized in the vicinity of atom and one may compute the local moments by integrating the density in a small region of space containing an atom. This comprises a pragmatic approach, which is not highly sensitive to the exact choice of domain defining a site (and it is the approach applied here).

\section{Discussion}\label{sec:discussion}

We have reported high-throughput DFT calculations of the lattice-mediated and clamped-ion ME response along with spin and orbital magnetization analogs in 124 FM and AFM monolayers.
This work fully characterizes the total ME response tensor in these compounds and our primary finding is that one cannot neglect the contributions from individual terms as they may add or separately dominate the response amplitude as a whole.
Furthermore, since the relative signs of the different contributions may differ, they can add up to yield a response that is much smaller than predicted by individual components, as shown in the case of $\mathrm{TaO}_2$.
We found that the electronic (spin) effect is most often the dominating contribution in the results, being the primary contribution in 50$\%$ of the non-zero components, and we have highlighted the FM ML $\mathrm{MnWSeTe}_3$ as an extreme example of this.
This is perhaps surprising since the LM spin term is often regarded as the primary contribution to static ME response.
%
%, due to the large responses induced by ionic motion under a field\cite{Bousquet2011}.
%
However, we should point out that other studies have shown that the CI,S contribution can be quite large.
For example, in the case of troilite FeS\cite{Ricci2016}, it was demonstrated that the CI,S contribution can comprise almost 60$\%$ of the total response. 

We observe that AFMs tend to display a larger response than FMs.
The anti-ME response was investigated using the local moment picture and it was shown that this effect can be significant in the orbital ME response as well as in the spin components and that it is also present in FMs as well as AFMs.
We should point out that materials predicted from theoretical databases may pose challenges with respect to experimental synthesization even if they are predicted to be stable by first-principle calculations.
Additionally, a single monolayer that has broken inversion symmetry may recover inversion in the lowest energy stacking configuration\cite{Pakdel2024} - thus removing the ME property.
Alternatively, bilayers composed of inversion symmetric monolayers may stack to remove inversion as in the case of $\mathrm{CrI}_3$ thus giving rise to an appreciable ME coupling\cite{Lei2021}.
Layer stacking (in particular sliding and twisting) remains an interesting degree of freedom for possible tuning the ME response\cite{Liu2020, Pakdel2024} and we leave this as an open direction for future studies.
Finally, we should comment on some of the limitations of this work.
We assumed collinear ground states and then computed one SCF cycle to converge the electronic density self-consistently with LDA+SOC.
A stricter tolerance may find different ground states (typical values could be ranging from a maximum absolute change in integrated density of $10^{-8}$ to $10^{-10}$  $\mathrm{e}^-$/valence electron), which would then alter the expected the ME tensor symmetry.
For our study, the magnetic moments usually stay in their imposed configurations but there are some cases where there can be large moment cantings.
This indicates that it is possible that the true ground state may exhibit a non-zero supercell ordering vector\cite{Sodequist2024}.
Such a helical or spiral order may impose a vanishing linear ${\bm \alpha}$\cite{Mangeri2025} or, in some cases, a spontaneous electronic polarization can arise from this spin texture\cite{Ruff2015} (type II multiferroic) which may enhance the ME effect.

We should also mention that proper treatment of electronic correlations may also be an influential factor in determining ${\bm \alpha}$.
For the case of $\mathrm{Cr}_2\mathrm{O}_3$, we have previously found that the application of Hubbard +U corrections can appreciably change the various contributions\cite{Mangeri2024}.
Lastly, the DFT calculations are carried out for $T = 0$ K and it is worth noting that many experiments exploring the ME effect are conducted at finite temperatures.
The methodology presented by Mostovoy \emph{et} \emph{al}\cite{Mostovoy2010} was able to include the temperature dependence of ${\bm \alpha}$, and for the case of $\mathrm{Cr}_2\mathrm{O}_3$, it was shown that some components can change sign and grow in magnitude at elevated temperatures.
We anticipate that the space of theoretically predicted stable 2D materials will continue to expand.
Our reported data set is not particularly large, which prevents us from revealing general physical trends across the materials.
Nevertheless, we hope that the present study may motivate further research in this direction.
In particular, it may be possible that machine learning could help to elucidate the important correlations between magnetoelectricity and other physical properties (i.e. magnetic network geometry, spin-phonon coupling, symmetric and antisymmetric exchange interactions, or ligand environments).

\begin{acknowledgments}

\vspace*{-10pt}

We thank M. Ovesen for helpful suggestions regarding convergence of the calculations.
The authors acknowledge funding from the Villum foundation Grant No.~00029378. 
Computations were performed on Nilfheim, a high performance computing cluster at the Technical University of Denmark.
We acknowledge support from the Novo Nordisk Foundation Data Science Research Infrastructure 2022 Grant: A high-performance computing infrastructure for data-driven research on sustainable energy materials, Grant no. NNF22OC0078009.

\end{acknowledgments}

\bibliography{apssamp}% Produces the bibliography via BibTeX.

\end{document}